\documentclass[aps,prc,showpacs]{revtex4}

\usepackage{epsfig}
\usepackage{dcolumn}
\usepackage{amsfonts}
\usepackage{color}

\newcommand{\be}{\begin{equation}}
\newcommand{\ee}{\end{equation}}
\newcommand{\ba}{\begin{array}}
\newcommand{\ea}{\end{array}}
\newcommand{\bea}{\begin{eqnarray}}
\newcommand{\eea}{\end{eqnarray}}
\def\hf{\textstyle{1\over2}}

\def\Rb{\mathbb{R}}

\begin{document}


\title{Construction of SU(3) irreps in canonical SO(3)-coupled bases}


\author{D.J.\ Rowe}
\affiliation{Department of Physics, University of Toronto,
Toronto, Ontario M5S 1A7, Canada}
\author{G.\ Thiamova}
\affiliation{Department of Applied Mathematics, University  of Waterloo, Waterloo,
Ontario N2L 3G1, Canada}

\affiliation{Nuclear Physics Institute, Czech Academy of Sciences,
Prague-Rez, 250 68 Czech Republic}


\begin{abstract}
 Alternative canonical methods for defining canonical SO(3)-coupled bases for SU(3) irreps are considered and compared.  It is shown that a basis that diagonalizes a particular linear combination of SO(3) invariants in the SU(3) universal enveloping algebra gives basis states that have good $K$ quantum numbers in the asymptotic rotor-model limit.

\end{abstract}

\maketitle

\section{Introduction}

A common problem in the construction of group or Lie algebra representations is to define a canonical basis in situations where multiplicities occur.
For example, bases which reduce the subgroup chain
\be \begin{array}{ccccc}
{\rm SU(3)} & \supset  & {\rm SO(3)} &\supset &{\rm SO(2)}  \\
  (\lambda \mu)& K  & L&&M \end{array}
\ee
 are indexed by the quantum numbers $(\lambda\mu)$, $L$ and $M$, of the respective groups SU(3), SO(3), and SO(2).
 However, an extra label $K$ is required to distinguish different irreps of SO(3) that occur in a given SU(3) irrep.
 This paper, is concerned with useful ways to define orthogonal sets of such SO(3) irreps.

In principle, multiple occurrence of subgroup irreps can be defined in any arbitrary way.  However, it is useful to have a well-defined ``canonical" definition, that can be reproduced by anyone so that results derived by one person are meaningful to someone else.
For example, in applications of group representations, considerable use is made of Clebsch-Gordan coupling and Racah recoupling coefficients that are defined for particular resolutions of the multiplicities that occur.  Two kinds of multiplicity arise:  one is the multiplicity in the choice of basis for each irrep.  Another, of equal importance, is the multiplicity of different irreps that occur in the decomposition of tensor products of irreps.
In this paper, we address the resolution of the first of these two multiplicities.

\section{SU(3) irreps and their asymptotic limits}

 There are several ways to construct SU(3) irreps in an SO(3)-coupled basis and derive the corresponding matrices representing elements  of the su(3) Lie algebra.  Note that we use upper case letters to denote a Lie group, e.g., SU(3), and lower case letters, e.g., su(3), for its Lie algebra.
The su(3) Lie algebra is spanned by five components of a quadrupole tensor ${\mathcal{Q}}$ and three components of an angular momentum ${\bf L}$.
As we discuss below, su(3) has an asymptotic limit in which it contracts to the Lie algebra, rot(3), of a rigid-rotor model. The latter Lie algebra is likewise spanned by five components of a quadrupole tensor ${\mathcal{Q}}$ and three components of an angular momentum ${\bf L}$.
Both su(3) and rot(3) have commutation relations
\bea &[ L_k, L_{k'}] = - \sqrt{2}\, (1k,1k'|1k+k')  L_{k+k'}  ,&\\
&[ L_k, {\mathcal{Q}}_\nu] = -\sqrt{6}\, (1k,2\nu |2 \nu+k)  {\mathcal{Q}}_{\nu+k}  .&\eea
However, they differ in the commutators of their $\{ {\mathcal{Q}}_\nu\}$ operators;
\be [ {\mathcal{Q}}_\nu, {\mathcal{Q}}_\mu] =3\sqrt{10}\, (2\mu,2\nu|1 \mu+\nu)
L_{\mu+\nu}\times
\cases{ 0 & for rot(3) ,\cr
1 &for su(3) .\cr}  \label{eq:su3.3}\ee
Thus, whereas su(3) is semi-simple, its contraction, rot(3), is a semi-direct sum of an abelian subalgebra, isomorphic to
$\Rb^5$, and an so(3) angular momentum algebra; we denote this by writing rot(3) $\simeq [\Rb^5] $so(3).

As shown in ref.\    \cite{JPE}, basis states for a generic $(\lambda,\mu)$,  irrep of su(3), are labeled by angular-momentum quantum numbers, $L$ and $M$, with $L$ running over the values
 \be L = \cases{ \lambda+K, \lambda+K-1, \ldots, K & for
$K\not= 0$\cr
\lambda, \lambda -2, \ldots , 0\; {\rm or}\; 1 & for
$K=0$}\label{eq:su3.35}\ee
with
\be K = \mu,\mu-2, \ldots  0\; {\rm or}\; 1 \, .\ee
Thus, in the generic case, there is a multiplicity of states with given values of $L$ and $M$, which can be indexed by $K$ or any other convenient label.

Irreps of the type $(\lambda,0)$ are particularly simple.
  They have orthonormal SO(3)-coupled bases given, without multiplicity, by a set of states
 \be \{ |LM\rangle ; M=-L, \dots,+L, \; L=\lambda, \lambda-2, \dots , 0 \;{\rm or}\; 1\},\ee
 in which $L$ runs over even or odd integer values according as $\lambda$ is, respectively, even or odd.  Reduced matrix elements for such multiplicity-free irreps have analytical expressions given \cite{RleBR,RVC} (in natural units) by the equations
 \bea &\langle L\|{ {\mathcal{Q}}}\| L\rangle =\sqrt{2L+1}\,
(L0,20|L0) (2\lambda+3),& \\
 &\langle L+2\|{ {\mathcal{Q}} }\| L\rangle = \sqrt{2L+1}\,
(L0,20|L+2,0)\, [4(\lambda-L)(\lambda +L+3)]^{\hf}  .& \eea

A systematic way to derive matrix elements for a generic SU(3) irrep was given  \cite{RleBR,RVC} in terms of vector coherent state \cite{VCS,HechtVCS} theory.
VCS methods were also used in a derivation of SU(3) Clebsch-Gordan coefficients in an SO(3)-coupled  basis \cite{RoweBahri, BRD}.
Conversely, a set of SU(3) Clebsch-Gordan coefficients computed in an SO(3)-coupled basis enables one to derive the SO(3)-reduced matrices of the SU(3) quadrupole tensor in that basis.  Examples of reduced matrix elements derived in this way are given below.
Such methods do not give analytical expressions for generic irreps, for which there are multiplicities,
However, analytical expressions are obtained  \cite{RleBR,RVC} in the asymptotic limits which are approached as either $\lambda$ or $\mu\to\infty$.

In the following, we restrict consideration to su(3) irreps $\{(\lambda,\mu)\}$ with $\lambda \geq \mu$.
This is because of the well-known fact (shown, for example, in \cite{RVC}) that the irreps
$(\lambda,\mu)$ and $(\mu,\lambda)$ are simply related.  Specifically, if
$\Gamma^{(\lambda\mu)}_\nu$ denotes the matrix representing the quadrupole operator
$\mathcal{Q}_\nu$ in the su(3) irrep $(\lambda,\mu)$, then
\be \Gamma^{(\lambda\mu)}_\nu = -\Gamma^{(\mu\lambda)}_\nu .\ee

With this restriction, asymptotic expressions for the su(3) quadrupole matrix elements are given for $\lambda \to\infty$ by
\bea \langle KL \| Q \| KL\rangle  &\sim& \sqrt{2L+1}\, \big[ (LK ,20|LK ) (\Lambda+\delta_{K1}\sigma_{LL})  \label{eq:AS1}
\eea
\bea
\langle KL+1 \| Q \| KL\rangle&\sim& \sqrt{2L+1}\,\big[ (LK ,20|L+1K )\sqrt{[(\Lambda-L-1+\delta_{K1}\sigma_{L+1,L})(\Lambda+L+1+\delta_{K1}\sigma_{L+1,L})]} \label{eq:AS2}
\eea
\bea
\langle KL+2 \| Q \| KL\rangle&\sim& \sqrt{2L+1}\,\big[ (LK ,20|L+2K)\sqrt{[(\Lambda-2L-3+\delta_{K1}\sigma_{L+2,L})(\Lambda+2L+3+\delta_{K1}\sigma_{L+2,L})]}
\eea
\bea
\langle K+ 2,L' \| Q \| KL\rangle &=& (-1)^{L'-L} \langle KL \| Q \| K+2,L'\rangle \nonumber\\
&\sim& \sqrt{(2L+1)(1+\delta_{K,0})}\,
(LK ,2, \pm 2|L',K\!\pm\!2)  \textstyle\sqrt{{3\over 2}(\mu - K )(\mu + K+2) } , \label{eq:AS3}
\eea
where $\Lambda=2\lambda+\mu +3$  and
\be \sigma_{L'L} = \hf (\mu+1) (-1)^{\lambda+L} \times
\cases{
 -{3L(L+1)\over 3-L(L+1)} & for $L'=L$ \cr
\quad L+1 &for $L'=L+1$ \cr
\quad -L &for $L'=L-1$ \cr
\quad -1 &for $L'=L\pm 2$  . } \label{eq:AS4} \ee

These asymptotic expressions are shown below to provide accurate approximate expressions for su(3) matrix elements for moderately large but finite values of $\lambda$.
They are similar in form to those of an irrep of the rot(3) rigid-rotor algebra), given  by
\bea \langle KL'\| \mathcal{Q} \| KL\rangle &=& \sqrt{2L+1}\, \big[ (LK,20|L'K)\, \bar q_0
\label{eq:rotor1}\nonumber\\
& &  + \delta_{K,1}\,  (-1)^{\lambda +L+1} (L,-1,22)L'1) \textstyle
 \bar q_2\big], \label{eq:10}\\
\langle K+2,L'\| \mathcal{Q} \| KL\rangle &=&  (-1)^{L'-L} \langle KL \| Q \| K+2,L'\rangle \nonumber\\
&=&\sqrt{(2L+1)(1+\delta_{K,0})} \, (LK,22|L',K+2)\, \bar q_2 , \label{eq:rotor2}
\eea
with
\be \bar q_0 = 2\lambda+\mu +3 , \quad \bar q_2  = \sqrt{\frac{3}{2}} \, \mu .\ee
The latter expressions give accurate approximations when both $\lambda$ and $\mu$ are large but are generally not as accurate as those of eqns.\ (\ref{eq:AS1}) - (\ref{eq:AS3}).

A computationally simple method \cite{GTW}, used in the present calculations, for deriving numerically precise matrix elements of an su(3) irrep  is to start from two known irreps, $(\lambda_1,0)$ and $(\lambda_2,0)$, and diagonalize the
SO(3)-invariant operator $ {\mathcal{Q}}\cdot  {\mathcal{Q}}$  in the tensor product of these irreps, where
$ {\mathcal{Q}}:=  {\mathcal{Q}}^{(1)}+ {\mathcal{Q}}^{(2)}$ is the summed quadrupole tensor for the two irreps.  To within a term proportional to the SO(3) Casimir invariant, ${\bf L}\cdot{\bf L}$,  the operator $ {\mathcal{Q}}\cdot  {\mathcal{Q}}$ is proportional to the SU(3) Casimir invariant.  Thus, its eigenstates belong to SU(3) irreps and, in the process of deriving them, one obtains all the reduced matrix elements of the quadrupole tensor (albeit in a basis chosen arbitrarily by the computer).
However, as shown in ref.\ \cite{GTW}, if one then diagonalizes the operator 
$ {\mathcal{Q}}^{(1)}\cdot{\mathcal{Q}}^{(2)}$ within the space of a 
 $(\lambda,\mu)$ irrep within the tensor product of $(\lambda_{1},0)$ and 
 $(\lambda_{2},0)$ irreps, then the degeneracies are lifted and the multiplicity of SO(3) irreps is resolved.
Simple techniques   for constructing such basis states and deriving their matrix elements were given in ref.\ \cite{GTW} and are used in the present calculations.  
Examples of reduced quadrupole matrix elements obtained in this way for the (32,5) and (10,4) irreps are shown in the columns of 
Tables \ref{table1} and \ref{table2} labeled  
$ {\mathcal{Q}}^{(1)}\cdot{\mathcal{Q}}^{(2)}$.
However, this bases does not appear to correspond to any of the canonical bases we consider below.

\section{Alternatives for resolving the SO(3) multiplicities}

 We consider three alternatives.

\subsection{Alternative I}

A standard way to resolve the SU(3) $\supset$ SO(3) multiplicity is by eigenstates of the  angular-momentum-zero coupled operator
\be X_3:=( {\bf L}\otimes  {\mathcal{Q}} \otimes{\bf L})_0. \ee
This  operator   is an SO(3) scalar in the SU(3) universal enveloping algebra \cite{Judd}.
Its potential use for resolving the SU(3) $\supset$ SO(3) multiplicity was noted by Bargmann and Moshinsky  \cite{BargM}. Such a use  is easily implemented because matrix elements of $X_3$ in any SO(3)-coupled basis for an SU(3) irrep are given to within an unimportant $L$-dependent constant, $c_L$,  by
\be  \langle \beta L' \| X_3  \| \alpha L\rangle =\delta_{L',L} c_L \langle \beta L \| {\mathcal{Q}} \| \alpha L\rangle.\ee
Thus, an SO(3)-coupled basis that diagonalizes
$X_3$ is given by the eigenstates of the $M^L$ matrices with elements
\be M^L_{\beta\alpha} := \langle \beta L \| {\mathcal{Q}} \| \alpha L\rangle .\ee
A variant of this method was used in the construction of bases for VCS irreps by 
$K$-matrix methods \cite{Kmat,U5}.
Examples of reduced quadrupole matrix elements in such a basis are given in Tables \ref{table1} and \ref{table2}.

\begin{table}[ht]
\caption{\label{table1} Comparison of quadrupole reduced matrix elements
$\langle K_fL_f \|\mathcal{Q}\| K_iL_i\rangle$ for bases defined by 
diagonalizing the operator $ {\mathcal{Q}}^{(1)}\cdot{\mathcal{Q}}^{(2)}$  and by the I, II and III alternatives, as defined in the text, for the SU(3) irrep $(32,5)$.  Values given by the asymptotic approximations of eqns. 11-15 are shown in the column headed A.S.
Values for rot(3), given by eqns. 16-17, are shown in the column headed ROT(3). Values obtained by the alternative algebraic approach described in Ref. \cite{GTW}, corresponding to the quadrupole-quadrupole interaction strength C=3.99, are in the last column.}

\begin{center}
\begin{tabular}{|c|c|c|c|c|c|c|c|} \hline
$K_i\; L_i$ & $K_f\; L_f$ &$ {\mathcal{Q}}^{(1)}\cdot{\mathcal{Q}}^{(2)}$& I & II & III & A.S. & ROT(3)  \\ \hline \hline
$1;3$ &$1;1$ 
  & 58.030319 &  81.421678 &  81.979149
  & 81.974076  & 81.975606  & 81.610661
\\

$3;3$ &$1;1$
& 59.131052& 15.313707  &  11.975740
  & 12.010416  & 12  &10.606602
  \\

$1;3$ &$1;2$ 
   & -71.091833 & -81.321756  & -80.940834
   & -80.945466   & -80.944425  & -79.5
\\

$3;3$ &$1;2$
   &-40.223759  & 7.666261   & 10.980970
   & 10.946736  & 10.954451  & 9.682458
\\

$1;3$ &$1;3$ 
  &  -1.248765 & -61.784810  &  -61.476192
  & -61.482530   & -61.481705   & -63.531095
\\

$3;3$ &$1;3$ 
  & -86.819713& 0  &  7.551035
  &  7.473012   & 7.483315  & 6.614378
\\

$3;3$ &$3;3$ 
  & 62.730470     &123.266515   &  122.957882
  & 122.964235   & 122.963409  &  122.9634092
\\

$1;4$ &$1;2$
  &   90.971036   &  99.475476 &  100.490358
  & 100.480593   & 100.484540 & 101.737583
\\

$3;4$ &$1;2$ 
   &  44.061450  &17.991295& 10.900948
   &  10.990508 & 10.954451 & 9.682458
\\

$1;4$ &$1;3$
  & -66.087883 &  -52.793962 &  -51.839707
  &  -51.849622  & -51.845926  & -51.693575
\\

$3;4$ &$1;3$
  &  69.315355 & 14.450569  &  12.953794
  & 12.961550  &12.961481  & 11.456439
\\

$1;4$ &$3;3$
  & -0.916687   &  3.009507 &  -3.8839466
  &   -3.792147 & -3.794733 & -3.354102
 \\

$3;4$ &$3;3$
  & -99.882581     & -127.061082  &  -127.590584
  & -127.588524   & -127.589968 &  -127.787323
\\

$1;4$ &$1;4$
  &  -97.415026   &  -108.055071 &  -107.316055
  &  -107.334482 &  -107.331699& -105.038286
\\

$3;4$ &$1;4$
  & -38.133331 & 0  &  10.407601
  & 10.277606  & 10.297396  & 9.101698
\\

$3;4$ &$3;4$
  &28.612663    & 39.252708  &  38.513682
  &  38.532119  & 38.529328 &  38.529328
\\

$1;5$ &$1;3$
  & 85.179826    & 126.274052  &  128.974225
  & 128.958781  &129.336770  & 129.037785
\\

$1;5$ &$3;3$
  & -0.269230 & -3.674329  &  1.317411
  & 1.274925  & 1.264911 &  1.118034
\\

$3;5$ &$1;3$
  &60.505815   &  26.583826 &  10.495438
  &  10.647742 &  10.583005 & 9.354143
\\

$3;5$ &$3;3$
  & 65.665141  &  69.189233 &  69.022209
  &  69.009663 & 69.229088 & 69.558608
\\

$5;5$ &$3;3$
  & -76.363826 &  16.305010 &  14.425467
  & 14.515278  & 14.491377 &  16.201852
\\

$1;5$ &$1;4$
  & -83.593639  & -69.460806  &  -67.269069
  &  -67.283818 & -67.278526  & -65.453419
\\

$1;5$ &$3;4$
  & 82.77925   & 9.338480  & - 5.939770
  &  -5.807912 & -5.796551 & - 5.123475
1\\

$3;5$ &$1;4$
  & -16.878049 & 13.723799  &  14.185688
  &  14.212501  & 14.198591 &12.549900
\\

$3;5$ &$3;4$
  & -68.306769   & -135.235541  &  -136.270512
  &  -136.278511 & -136.280593 & -136.610395
\\

$5;5$ &$3;4$
  &  68.027894 & 5.913498  &  9.628153
  &  9.450810   & 9.486833  & 10.606602
\\

$1;5$ &$1;5$
  & -48.716838   &  -95.531881 &  -93.682063
  &  -93.716833  & -93.712654  & -96.231811
\\

$3;5$ &$3;5$
  & 43.126580   & -10.404218  &  -12.110821
  & -12.089411  & -12.091955  & -12.091955
\\

$5;5$ &$5;5$
  & 81.164960    &  181.510801 &  181.367616
  &  181.380946 & 181.379330  &  181.379330
\\

$3;5$ &$1;5$
  &  -52.087438 & 0                &  12.414743
  &  12.299948  & 12.313845 &10.884004
\\

$5;5$ &$3;5$
  & 113.765385 &  0                &  5.258647
  &  5.006770    & 5.038315    &  5.633007
\\

\hline
\end{tabular}
\end{center}
\end{table}

\begin{table}[ht]
\caption{\label{table2} Comparisons of quadrupole reduced matrix elements
$\langle K_fL_f \|\mathcal{Q}\| K_iL_i\rangle$ as described in Table \ref{table1} for the SU(3) irrep $(10,4)$.}

\begin{center}
\begin{tabular}{|c|c|c|c|c|c|c|c|} \hline
$K_i\; L_i$ & $K_f\; L_f$ &$ {\mathcal{Q}}^{(1)}\cdot{\mathcal{Q}}^{(2)}$& I & II & III & A.S. & ROT(3)\\\hline \hline
$0;2$ &$0;0$ 
   &  19.146198  & 25.227104   & 26.854801
   & 26.823096   & 26.832816   &  27
\\

$2;2$ &$0;0$
  & 20.625775 & 12.473680  &  8.415442
  & 8.515925   & 8.485281    &  6.928203
\\

$0;2$ &$0;2$
   & -25.280167   & -33.827282   & -32.203990
   & -32.280594   & -32.271172   &  -32.271172
\\

$2;2$ &$0;2$
   & -22.476614 &  0                 & 10.353197
   & 10.111788  & 10.141851   &  8.280787
\\

$2;3$ &$0;2$
  & 32.612214  & 19.722620  &  13.305980
  & 13.464860  & 13.416407  &  10.954451
\\

$2;3$ &$2;2$
  & -30.272798  & -39.887555  &  -42.461171
  & -42.411040  & -42.426407  & -42.690748
\\

$0;4$ &$2;3$
  & 20.939094  & 10.108085  &  -9.182851
  & -8.552679  & -8.485281   & -6.928203
\\

$2;4$ &$2;3$
  &  -32.438270 & -41.477548  &  -41.183272
  & -41.375935  & -41.366653  &  -41.828220
\\

$4;4$ &$2;3$
  &  -18.966068 & 5.276214  &  8.367394
  & 8.079759     & 8.197561  &  8.197561
\\

$0;4$ &$0;2$
   & 29.639536 &  34.848900 &  41.886357
  & 41.742972  &  41.815923 &  43.296321
\\

$0;4$ &$2;2$
  & 0.058923  & -3.976255  &  2.594292
  & 2.343090  & 2.267787   &  1.851640
\\

$2;4$ &$0;2$
   & 19.653177 & 22.023088  &  8.545059
  & 9.080523    & 8.783101    & 7.171372
\\

$2;4$ &$2;2$
  & 21.468704 & 28.265009  &  26.965058
  & 26.958674 & 26.992062  &  27.947655
\\

$4;4$ &$2;2$
  & 30.379909  & 12.821131 &  10.867348
  &  11.055995 & 10.954451 &  10.954451
\\

$0;4$ &$0;4$
  & -45.391345  &  -48.446517 &  -40.877618
  &  -41.340626 &  -41.281422 &  -41.281423
\\

$2;4$ &$2;4$
   & 11.845141   &  -9.673158  & -16.863115
   &  -16.470046 & -16.512569 & -16.512569
\\

$4;4$ &$4;4$
 & 33.546210  & 62.755119  &   57.740733
 &  57.810677 & 57.793992  &   57.793992
\\

$2;4$ &$0;4$
  & -11.594159 &   0              &  15.396745
  &  15.025873 & 15.073844 &  12.307742
\\

$4;4$ &$2;4$
   &-33.333631 & 0               & 5.247793
   &  4.721849  & 4.854239  & 4.854239
\\

\hline
\end{tabular}
\end{center}
\end{table}

 \subsection{Alternative II}

A second alternative is to use generally accepted SU(3) Clebsch-Gordan coefficients  in an SO(3)-coupled basis to derive reduced matrix elements of the SU(3) quadrupole operator   by means of the identity
\be \langle \beta L' \| {\mathcal{Q}} \| \alpha L\rangle = \Big[\frac43 (2L'+1) (\lambda^2 + \mu^2 + \lambda\mu + 3\lambda +3\mu)\Big]^{\frac12} \big( (\lambda\mu)\alpha L; (11)2 \| (\lambda\mu) \beta L'\big) ,  \label{eq:alt2}
\ee
where  $(\lambda^2 + \mu^2 + \lambda\mu + 3\lambda +3\mu)$ is proportional to the value of the SU(3) Casimir operator for the $(\lambda,\mu)$  irrep and
$\big( (\lambda\mu)\alpha L; (11)2 \| (\lambda\mu) \beta L'\big) $ is an SO(3)-reduced SU(3)
Clebsch-Gordan coefficient.

In principle, the resolution of the SU(3) $\supset$ SO(3) multiplicity, defined in this way, is only canonical to the extent that the Clebsch-Gordan coefficients are themselves expressed relative to  a canonical basis.
But, even if they are not, provided they are freely available, they serve the practical purpose of making it possible to compare the results of calculations by different researchers who use a common set of such coefficients.  For present purposes, we use the Clebsch-Gordan coefficients
of refs.\ \cite{RoweBahri, BRD}. Some results are shown for comparison with the other alternatives in Tables  \ref{table1} and \ref{table2}.
The comparisons show a remarkable similarity between these results and those of the following alternative.  This will be explained in the Discussion.

 \subsection{Alternative III}

A third alternative is given by basis states which diagonalize a specified linear combination of the SO(3) scalar operators $X_3$ and
\be X_4:= ( {\bf L}\otimes [ {\mathcal{Q}}\otimes {\mathcal{Q}}]_2 \otimes{\bf L})_0\ee
 within the space of an SU(3) irrep.

The rationale for choosing a particular linear combination is based on the observation that there is a natural resolution of the SO(3) $\subset$ SU(3) multiplicity in the contraction limit in which an irrep of the su(3) algebra  progresses asymptotically towards an irrep of the rotor model algebra, denoted rot(3).
In particular, as pointed out in ref.\ \cite{RR},  the  intrinsic quadrupole moments of a
 rot(3) irrep, for which there is a naturally-defined  SO(3)-coupled basis, are related to the  SO(3) invariants   $\bar X_3$ and  $\bar X_4$, where the latter operators are defined, as for  the corresponding su(3) operators $X_3$ and $X_4$, but in terms of the commuting rot(3) quadrupole operators.
Because of the understood contraction of su(3) $\to$ rot(3) for large values of 
$\lambda$,
these observations suggested similar relationships for su(3).
Further relationships between the  rigid-rotor model and the SU(3) model were developed  by Leschber and Draayer \cite{LD}.

The su(3) $\to$ rot(3) contraction is derived as follows \cite{contract}. Let
\be \epsilon(\lambda\mu) := \frac12
 \Big[\lambda^2 + \mu^2 + \lambda\mu + 3\lambda +3\mu\Big]^{-\frac12}  ,\ee
  denote the inverse square root of the eigenvalue of the SU(3) Casimir invariant
\be C_2 := \mathcal{Q}\cdot \mathcal{Q} + 3{\bf L}\cdot {\bf L} \ee
  for the irrep $(\lambda,\mu)$,
 and let $Q$ denote the su(3) quadrupole tensor in inverse units of $\epsilon(\lambda\mu)$, i.e.\ the tensor with components
\be Q_\nu :=  \epsilon(\lambda\mu) \mathcal{Q}_\nu .\ee
It follows from eqn.\ (\ref{eq:alt2}) that, for values of $L\ll \lambda$,  the non-zero reduced matrix elements of $Q$
  are of order of magnitude
 \be \langle \beta L' \| Q \| \alpha L\rangle \lesssim \textstyle
  \big[ \frac43 (2L'+1)\big]^{\frac12} .\ee
 Moreover, the rhs of the commutation relation
\be [ Q_\nu, Q_\mu] =3\sqrt{10}\, (2\mu,2\nu|1 \mu+\nu)\,\epsilon(\lambda\mu)^2\,
L_{\mu+\nu} , \label{eq:su3cr}\ee
becomes negligible when used with states of angular momentum $L$ for which
\be \epsilon(\lambda\mu)^2  L  \ll 1.  \label{eq:e2L}\ee
Thus, within the subspace of states of angular momentum $L$ for which eqn.\ (\ref{eq:e2L}) is satisfied, the matrix elements of an su(3) irrep become indistinguishable from those of a rot(3) irrep.
In this situation, su(3) is said to contract to rot(3).

This contraction  is of considerable interest in nuclear physics for explaining the origins of rotational structure in terms of the nuclear shell model in an SU(3) $\supset$ SO(3) coupled basis.
Because there is a natural resolution of the SO(3) $\subset$ ROT(3) multiplicity, the su(3) $\to$ rot(3) contraction, is also significant for the resolution of the SO(3) $\subset$ SU(3) mutiplicity.

The results of ref.\ \cite{RR} suggest that the above defined basis states of the rot(3) algebra should diagonalize a linear combination of the $\bar X_3$ and  $\bar X_4$ operators.
To ascertain that this is true and determine the linear combination, we consider the ratios of the matrix elements
\be  R(L,K) := \frac{
\langle K+2, L\|  \bar X_4\|KL\rangle} {\langle K+2, L\|  \bar X_3\| KL\rangle} = \frac
{\langle K+2, L\|   [ {\mathcal{Q}}\otimes {\mathcal{Q}}]_2  \| KL\rangle}
{\langle K+2, L\|  {\mathcal{Q}} \| KL\rangle} \ee
for  rot(3)  irreps.
Equations (\ref{eq:rotor1}) and (\ref{eq:rotor2}) give the reduced rot(3) matrix elements
$\langle K L\|  {\mathcal{Q}} \| KL\rangle $ and
$\langle K+2, L\|  {\mathcal{Q}} \| KL\rangle$.
From them, we can evaluate
\be  \langle K+2, L\|  \big[ \mathcal{Q}\otimes\mathcal{Q}\big]_2 \|KL\rangle
= \sum_{K_1L_1} U(L2L2;L_12)
\frac{\langle K+2, L\|  {\mathcal{Q}} \| K_1L_1\rangle  \langle K_1L_1\|  {\mathcal{Q}} \| KL\rangle}
{\sqrt{2L_1+1}} \ee
and the ratio $R(KL)$ for any values of $L$ and $K$.  In this way, it is determined that
$R(LK)$ takes the $L$- and $K$-independent value
\be R(LK) = \sqrt{\frac{8}{7}} \,\bar q_0 .\ee
This result means that the basis states of the rigid-rotor rot(3) algebra with good $K$ quantum numbers are eigenstates of the SO(3)-invariant
\be \bar Z :=\bar X_4 - \sqrt{\frac{8}{7}} \,\bar q_0 \bar X_3 .\ee

Similarly, we can define  basis states for an SU(3) irrep to be eigenstates of the corresponding
SO(3)-invariant
\be Z := X_4 - \sqrt{\frac{8}{7}} \,(2\lambda + \mu +3) X_3 , \label{eq:Z}\ee
with the expectation that, in such a basis, the su(3) quadrupole matrix elements between states of $L\ll \lambda$ will aproach those of a rot(3) irrep in the asymptotic limit.
Such basis states are uniquely defined, and provide a physically relevant resolution of the SO(3) multiplicity for any SU(3) irrep.
Results obtained for such SU(3) bases are shown in Tables  \ref{table1} and \ref{table2}.

\section{Discussion}

Tables  \ref{table1} and \ref{table2}  show comparisons of reduced quadrupole matrix elements obtained for the  alternatives given above for defining orthonormal SO(3)-coupled basis states for SU(3) irreps.
The tables also show the corresponding results given by the asymptotic approximation of eqns.\
(\ref{eq:AS1}) - (\ref{eq:AS4}) and for the rot(3) matrix elements given by  eqn.\ (\ref{eq:rotor1}) and (\ref{eq:rotor2}).
It should be emphasized that the results given for the SU(3) matrix elements listed in the columns headed I, II, and III are all numerically accurate to the precision shown; they only differ to the extent that they were computed relative to different bases.
The asymptotic results in the column headed A.S.\   are expected to agree with those of column III for values of $\lambda \gg L_i$.  Those listed in the column headed ROT(3) are for the rot(3) rotor algebra and likewise are expected to approximate those of columns III and A.S.\ when both $\lambda \gg L_i$ and $\mu \gg L_i$.

It can be seen that alternatives II and III are the same to within 1-3\% and, as expected, both  are consistent with an approach to the results of the asymptotic limit for large values of $\lambda$.
The results of diagonalizing the SO(3) invariant, $X_3$, in alternative I are qualitative similar but it is clear that the eigenstates of the linear combination of $X_3$ and $X_4$, given by $Z$ in eqn.\ (\ref{eq:Z}), give results much closer to those of asymptotic rotor-model limit.
The equivalence of results II and III is quite remarkable and fortuitous in the sense that it means that the bases used in the calculation of SU(3) $\supset$ SO(3) Clebsch-Gordan coefficients in refs.\  \cite{RoweBahri, BRD} can, in fact, be regarded as canonical in the above-defined sense that the basis states are eigenstates of a Hermitian operator.
This result was unexpected because the choice of basis states for an SU(3) irrep used in the computation of the SU(3) Clebsch-Gordan coefficients given in refs.\  \cite{RoweBahri, BRD}, did not make use of the SO(3)-invariant operator, $Z$.  However, the construction of the basis states that were used did make use of rotor-model methods  which, in the asymptotic limit likewise give standard rot(3) results.  Thus, in retrospect, it is understood that the Clebsch-Gordan coefficients obtained should be consistent with the SU(3) bases states defined by alternative II.

The tabulated matrix elements also show the expected result that the accurate matrix elements for the basis III are given more accurately, for small values of $\mu$  by the asymptotic SU(3) results of column A.S.\ than by those of the ROT(3) limit.

In conclusion, we remark that the above results provide a physical and practical resolution of the so-called inner, i.e., SU(3) $\supset$ SO(3), multiplicity problem.  However, the outer multiplicity that occurs in the decomposition of tensor products of SU(3) irreps is also of importance and, at present, we know of no canonical way to resolve it.

\acknowledgements
The authors are appreciative of helpful suggestions from G.\ Rosensteel.
DJR acknowledges financial support for his research from the Natural Sciences and Engineering Research Council of Canada.

\end{document}